\documentclass[aps,prl,twocolumn,superscriptaddress,showpacs,floatfix]{revtex4-2}

\usepackage{graphicx}% Include figure files
\usepackage{dcolumn}% Align table columns on decimal point

\usepackage{bm}% bold math
\usepackage{ulem}
\usepackage{soul}
\usepackage{color}
\usepackage{array}
\usepackage{multirow, makecell}
\usepackage{physics}
\usepackage{lineno}
\usepackage[english]{babel}

\usepackage[colorlinks,breaklinks,bookmarks=true,citecolor=blue,linkcolor=blue,urlcolor=blue]{hyperref}

 \newcommand{\khn}[1]{{\color{red}#1}}

\begin{document}

\preprint{APS/123-QED}

\title{%A tale of for a talk it is O.K. but I am not sure without reference to Dickens whether the title sounds to fairy-tale-ish
Two distinct gap structures in the mid-infrared
optical conductivity of the Hubbard model}

\author{
Dongwook Kim\textsuperscript{1,*} 
and Karsten Held\textsuperscript{1,*}
}

\affiliation{\textsuperscript{1}Institute of Solid State Physics, TU Wien, 1040 Vienna, Austria}

\email{dong.kim@tuwien.ac.at}
\email{held@ifp.tuwien.ac.at} % Corresponding author email

\date{\today}% It is always \today, today,

\begin{abstract}
Mid-infrared (MIR) optical conductivities in cuprate superconductors show universal features. In this study,  we demonstrate that these originate from two distinct gaps: the pseudogap mediated by antiferromagnetic spin fluctuations and the Mott-Hubbard gap. The MIR spectra from these two gaps show a characteristic and distinct shape and doping dependence, and can thus be distinguished. Specifically, 
the Mott-Hubbard correlations, while
 rendering a peak in the visible spectrum for small dopings, yield a broad optical response starting in the MIR when a ``waterfall'' develops in the one-particle spectrum at larger dopings. The numerically and analytically determined doping-dependence of both gaps consistently reproduce experiments, and provide a unified microscopic understanding of the cuprate MIR optical conductivity.  
\end{abstract}

\maketitle

\textit{Introduction --} Infrared (IR) peaks in the optical conductivity are universally observed in various cuprate superconductors~\cite{uchida1991optical, cooper1993optical, onose2001doping, basov2005electrodynamics, kumar2023characterization, kim2026optical} which all show remarkably similar systematic trends: In the underdoped regime, a prominent peak appears in the mid-IR (MIR) range, while a broad maximum is observed from the near-IR (NIR) to the visible (VIS) spectrum. Upon further hole doping, the MIR peak red-shifts. The NIR-to-VIS peak also red-shifts and evolves into a flat plateau ranging from MIR to NIR. Despite  this universally observed phenomenology, the origin of these peaks remains under debate.

Historically, the IR optical spectra of cuprate superconductors have been interpreted in terms of one-component and multi-component descriptions~\cite{basov2005electrodynamics}. The former attributes the Drude response and MIR continuum to a single itinerant electronic fluid with a strongly frequency-dependent scattering rate and effective mass.
%often associated with coupling to bosonic excitations~\cite{puchkov1996pseudogap}, 
Conversely, the latter phenomenology separates the spectrum into a coherent part and additional finite-frequency absorption channels, often interpreted as interband contributions or contributions from coexisting itinerant and localized carriers or from hopping carriers arising from strong electronic correlations~\cite{varshney1998two}. 
Both phenomenologies can reproduce selected features of the optical spectra, 
with the  one-component description being naturally more suitable for the Drude peak and adjacent MIR spectrum and the multi-component description being better tailored for describing separated MIR peaks. 
These two phenomenologies  need not be mutually exclusive at the microscopic level  and leave substantial freedom regarding the physical origin of the MIR spectra,  with conjectures ranging from polarons to impurities~\cite{basov2005electrodynamics} to Mott-Hubbard gaps~\cite{Jarrell1995} and the pseudogap (PG)~\cite{puchkov1996pseudogap}. 

\begin{figure*}[tb]
    \centering
    \includegraphics[width=.8\linewidth]{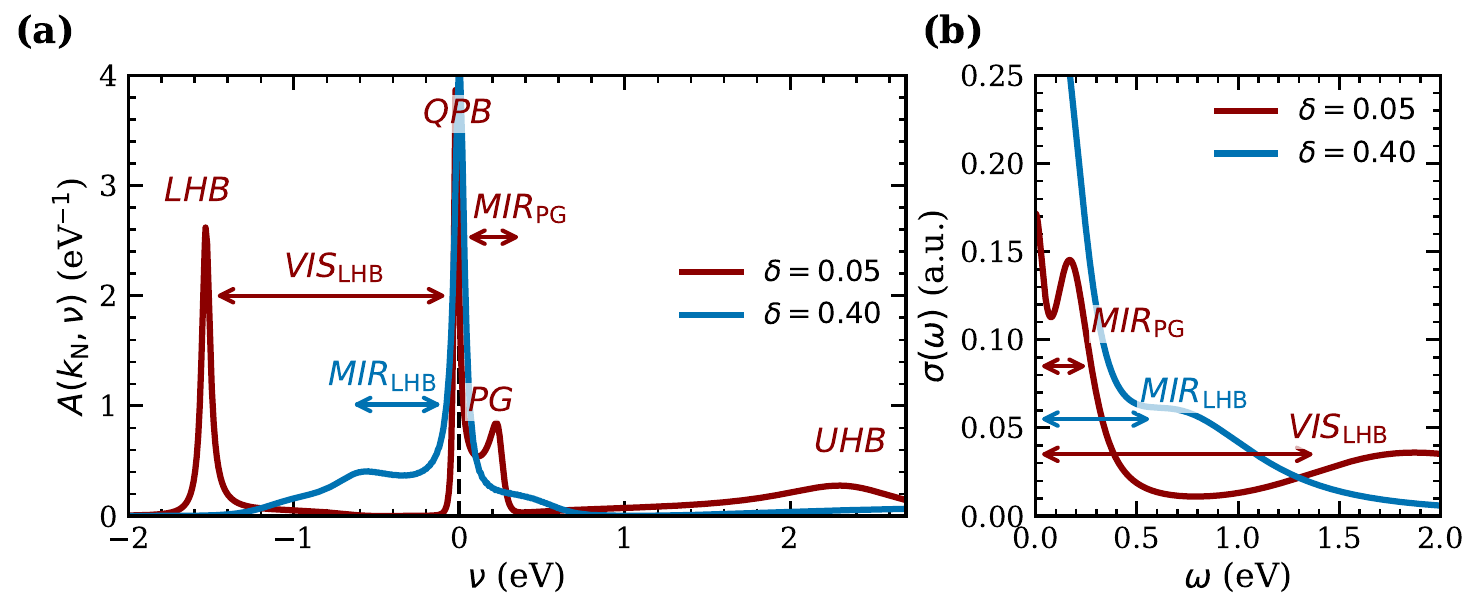}

    \caption{(a) Two different types of gaps in the nodal D$\Gamma$A spectral function. (b) The contribution to the optical conductivity in the MIR from optical transitions across the pseudogap (PG) and across the gap between lower Hubbard band (LHB) and quasiparticle band (QPB). For small doping the latter is in the visible spectrum.}
    \label{fig0}
\end{figure*}

In this paper, we elucidate the origin of the IR optical conductivity and its doping-dependence 
based on the ladder dynamical vertex approximation ($\mathrm{D}\Gamma\mathrm{A}$)~\cite{Toschi2007,RMPVertex} for the  Hubbard model on a square lattice.
Our findings establish a unified picture for the mid- and near-IR optical spectra in high-temperature superconductors, showing that they originate from two distinct gaps tied to  strong local correlation and bosonic spin fluctuations, respectively. 

The  description of the Mott-Hubbard gap is one of the cornerstones of dynamical mean-field theory (DMFT)~\cite{georges1996dynamical}, and corresponding optical spectral weight in the NIR-to-VIS has been reported in DMFT~\cite{Jarrell1995,Deng2013,Dasari2026, kang2021optical}.
Here we show that, with the development of a ``waterfall'', the Mott-Hubbard-related optical spectrum softens and broadens so that it eventually contributes to the MIR optical conductivity.
A spin-fluctuation-mediated PG, on the other hand, has been considered  
in the form of an unusual scattering rate
\cite{puchkov1996pseudogap,Tajima2016}, but 
can also lead to a hump~\cite{Munzar1999,Bergeron2011} or peak \cite{chakraborty2008spectral,lin2009optical} in the MIR range.
 By employing a non-perturbative microscopic description alongside a semi-analytical understanding, we reconcile and clearly distinguish the signatures of these two distinct gaps, which are intertwined with each other and the Drude peak, further complicating the analysis of the optical conductivity.

Our main finding is summarized in
 Fig.~\ref{fig0}:
In the doped Hubbard model, 
strong local correlations  split
the spectrum into a lower Hubbard band (LHB),
a quasiparticle band (QPB) and an upper Hubbard band (UHB). Low energy excitations within the QPB generate a Drude peak, see Fig.~\ref{fig0}\,(b).
Additionally, there are QPB-LHB and UHB-QPB
particle-hole excitations. At low dopings [red lines in Fig.~\ref{fig0}\,(a,b)], these excitations lie in the visible (VIS) range. Additional LHB-UHB optical excitations are separated by the Coulomb interaction $U$ and are the ultraviolet (UV). At larger hole-dopings [blue lines], a waterfall connects
LHB and QPB warping the dispersion such that the corresponding 1-particle spectra and optical excitations form a relatively broad peak from MIR to  NIR (MIR$_{\mathrm{LHB}}$).
%The (Mott-Hubbard) gap between QPB-LHB thus gives
%rise to a first optical peak in the (M)IR.

A second MIR peak (MIR$_{\mathrm{PG}}$) originates from the PG which is induced here by 
antiferromagnetic spin fluctuations. While the antinodal spectrum is gapped and forms Fermi arcs~\cite{norman1998destruction}, the nodal spectral function, which dominates  the optical conductivity, is  gapped but only  above the Fermi energy (at frequency $\nu=0$)
 [Fig.~\ref{fig0}\,(a)]. Consequently, in the underdoped regime, the nodal spectral function yields simultaneously a Drude peak and  an additional MIR peak in the optical conductivity via transitions across the PG [Fig.~\ref{fig0}\,(b)].

In the following, we systematically analyze the evolution of these two gaps through the pole structure of the self-energy. To this end, we investigate the square lattice Hubbard model
using the ladder D$\Gamma$A with $\lambda$ correction \cite{RMPVertex} at interaction
 $U=8t$ (corresponing to $3.2\,$eV for a typical hopping value of  $t=0.4\,$eV for cuprates), next-nearest neighbor and next-next-nearest neighbor hopping $t'=-0.25t$ and $t''=0.12t$, respectively, and $T=0.0167t$ (77\,K). The optical conductivity is calculated in the Kubo formalism from the D$\Gamma$A current-current bubble susceptibility.
For details on the model and method,
we refer the reader to the Supplementary Section 1 \cite{SM}. 
%For the detailed discussion, we first start with AFM-driven one.

\begin{figure}
    \centering
    \includegraphics[width=1.0\linewidth]{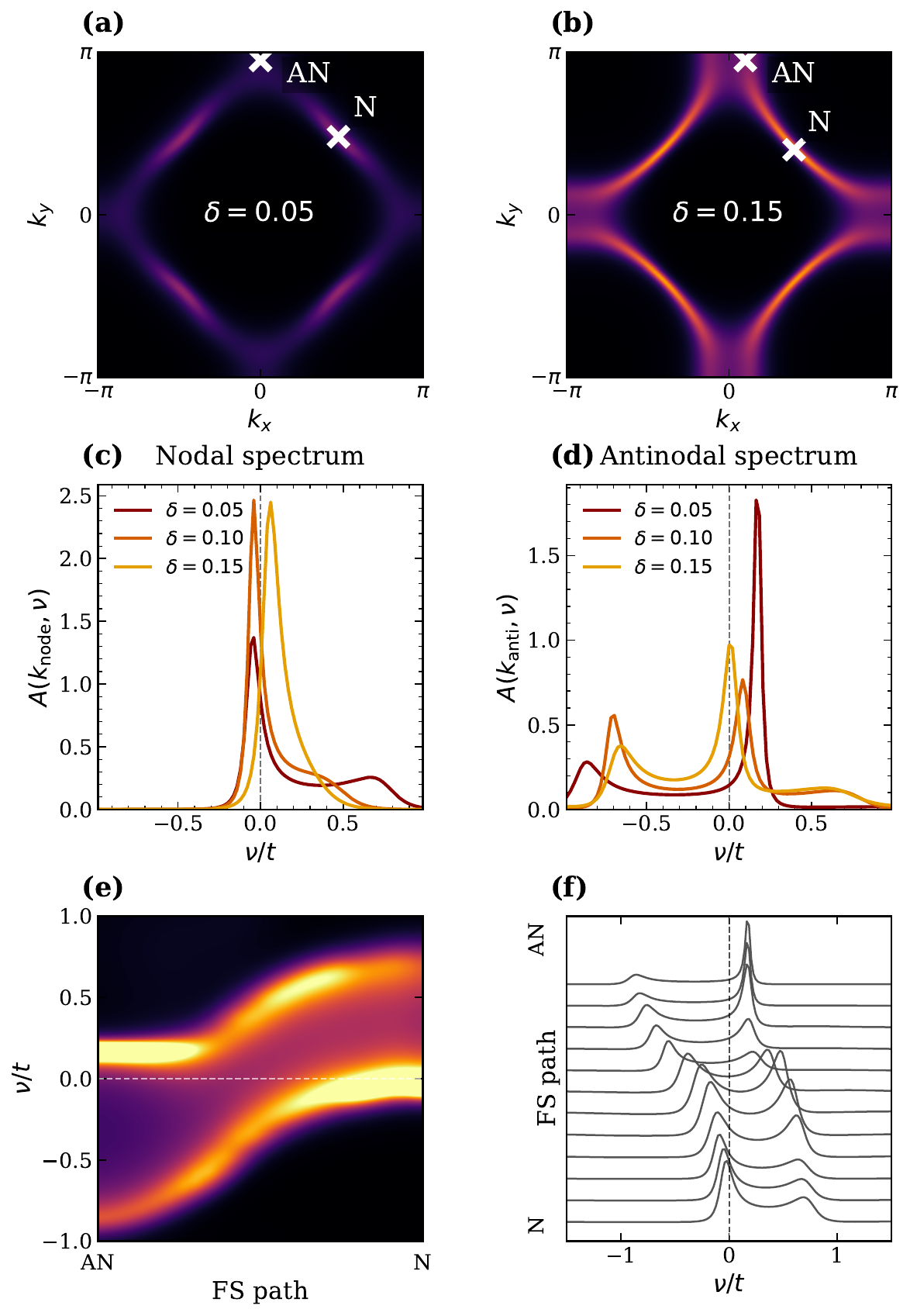}
    \caption{(a,b) Fermi surface obtained from $-\mathrm{Im}G(k, i\pi T)/\pi$ for (a) $\delta=0.05$ and (b) $\delta=0.15$. (c,d) Spectral function in the low-frequency range for  (c) nodal and (d) antinodal momentum in the underdoped regime. (e,f) Spectral function $A(k,\nu)$ along the Fermi surface (arc) at $\delta=0.05$ as (e) a heat map and (f) as a function of $\nu$ for a series of momenta along the arc.}
    \label{fig1}
\end{figure}

\textit{Pseudogap --} 
 Fig.~\ref{fig1}\,(a) shows the Fermi arc in the PG regime as obtained in D$\Gamma$A.
 Clearly, there is spectral weight along the nodal direction {$(k_x, k_y \!=\! k_x)$}, whereas there is no weight 
 along the antinode {$(k_x, k_y = \pi)$}. This is the very essence of the Fermi arc.
 Upon increase of the hole doping, the full Fermi surface is recovered, see
 Fig.~\ref{fig1}\,(b,d).
 
The gap structure also develops at the node.
However, as shown in Fig~\ref{fig1}\,(c), the gap at the node is above the Fermi energy with a persistent quasiparticle peak at the Fermi level and the size of the gap is slightly smaller than that of the antinode. At hole doping  $\delta = 0.15$, the gap closes at the node, 
 while there is still one at the antinode
 below the Fermi energy in Fig~\ref{fig1}\,(d).

 As shown in Fig.~\ref{fig1}\,(e) and (f), for low dopings, the gap never closes between antinode (AN) and node (N). But, its position shifts from below the Fermi energy (also including the Fermi energy) at the antinode to above the Fermi energy at the node. Only because the peak below this gap is always at the Fermi level for the node, we get a spectral weight at the node and a Fermi arc.
 
  This gap structure is of an anisotropic s-wave type. As the gap is above the Fermi energy at the node, angle-resolved photoemission spectroscopy (ARPES) experiments cannot observe it, as also pointed out in previous studies based on cellular dynamical mean-field theory (CDMFT)~\cite{stanescu2006fermi, sakai2010doped,sakai2013raman} and dynamical cluster approximation (DCA)~\cite{lin2010physics}.

\begin{figure}
    \centering
    \includegraphics[width=0.96\linewidth]{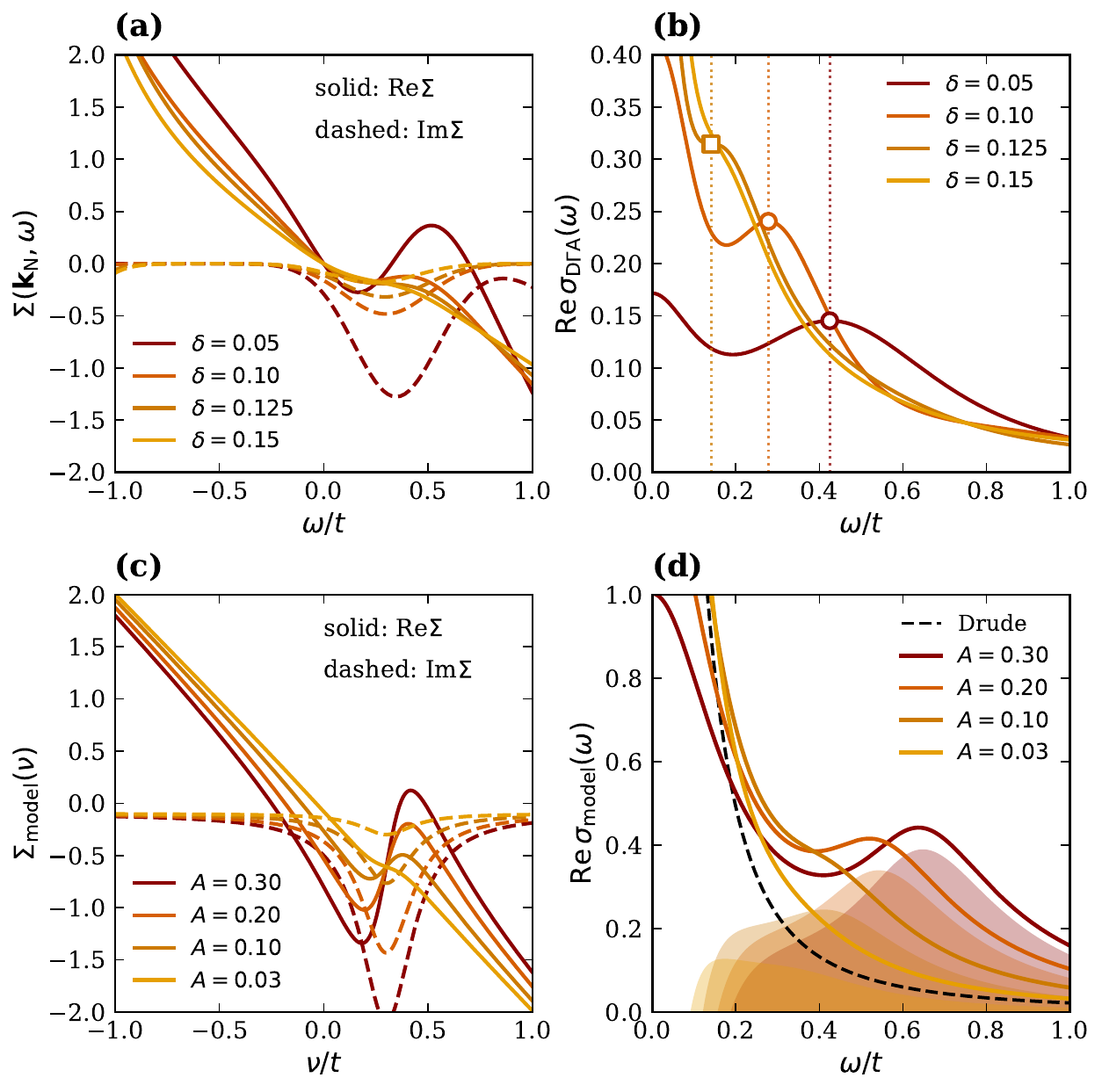}
    \caption{(a) Evolution of $\mathrm{Im}\Sigma(\omega)$ and $\mathrm{Re}\Sigma(\omega)$ at the node with doping. For better comparison, $\mathrm{Re}\Sigma(\omega)$ has been shifted so that $\mathrm{Re}\Sigma(0) = 0$ for all curves. (b) Optical conductivity for the same  dopings. Dotted vertical lines denote the peak frequency (c,d) 
    Same as (a,b) but for the model self-energy Eq.~(\ref{selfenergy_model}) for different weights $A$ of the pole. The black dashed curve in (d) corresponds to the Drude optical conductivity $\sigma_{Drude}(\omega)$ Eq.~(\ref{opt_drude}); and color fills represent the difference to the Drude optical conductivity, $\sigma(\omega) - \sigma_{Drude}(\omega)$.}
    \label{fig2}
\end{figure}

Due to the much larger Fermi velocity  near the node \footnote{Note that the energy-momentum dispersion is $\epsilon(\mathbf k)=-2t\big( \cos(k_x) + \cos(k_y)) -4t'\big( \cos(k_x)\cos(k_y )) - -2t''\big( \cos(2k_x) + \cos(2k_y))$ so that the Fermi velocity ${\mathbf v}(\mathbf k)=\partial \epsilon(\mathbf k)/\partial \mathbf k$ at antinodal momenta $(\pm \pi,0)$, $(0,\pm \pi)$ vanishes.}, it dominates the  optical conductivity. Fig~\ref{fig2}\,(a) shows the nodal self-energy, which has a pole structure of the form $1/(\nu-\nu_g+i\Delta)$ around $\nu_g\approx0.3t$. The broadening $\Delta$ cuts off the pole of $\Sigma$, and we get a finite Lorentzian maximum in $-{\mathrm{Im}} \Sigma$.
This (broadened) pole
opens the gap  in Fig~\ref{fig1}\,(c).
With increasing doping, the weight of the pole structure shrinks, and eventually vanishes for $\delta \approx 0.15$, where we have the closure of the nodal gap structure in Fig~\ref{fig1}\,(c). Fig.~\ref{fig2}\,(b) shows the 
MIR optical peak for the corresponding dopings, where the increase of doping red shifts MIR peak in the optical conductivity and eventually merges the MIR peak into the Drude peak at $\delta = 0.15$ as the nodal PG closes.

To attribute the MIR peak shift to the gap structure evolution in a controlled environment, we do a 
%The MIR peak frequencies also corresponded well to the size of the gap structure, while being slightly smaller than the peak distance in the nodal DOS due to the broad gap structure as shown in Fig.~\ref{fig1} (c). Thereby we can deduce the direct relevance of the AFM-driven nodal gap structure to the MIR peak. To concretize the idea that the gap structure \textcolor{blue}{\sout{corresponds to the}originates} MIR peak structure, we do a 
(semi-) analytical analysis using a model self-energy
\begin{equation}
\label{selfenergy_model}
\Sigma(\nu) = -i\Gamma_0 + \bigg(1-\frac{1}{Z}\bigg)\nu + \frac{A}{\nu - \nu_{g} + i\Delta} %% + \sum_{i}\frac{A_i}{\nu - \nu_{i} + i\Delta_i}
\end{equation}
where besides a gap structure with weight $A$
we also include a quasiparticle renormalization $Z$ for describing the behavior at low (real) frequency $\nu$ without the PG structure, and a constant scattering rate $\Gamma_0$.

For a linear bare dispersion and constant density of states, the optical conductivity without pole ($A=0$) is given by the  Drude formula
\begin{equation}
\label{opt_drude}
\sigma(\omega) \propto \frac{1}{\pi}\frac{2Z\Gamma_0}{\omega^2 + (2Z\Gamma_0)^2}
\end{equation}
where all the coefficients including the electronic charge and the bare Fermi velocity are included in the prefactor, see Supplementary Section 2~\cite{SM}. Thus, any non-Drude optical spectra in this model originates from the self-energy pole structure.

For the model analysis, we fix the parameters as $Z = 0.33$, $\Gamma_0 = 0.1$, $\nu_g = 0.3$ and $\Delta = 0.15$ to represent the nodal PG structure. The resulting model self-energy in Fig.~\ref{fig2}\,(c) roughly mimics the D$\Gamma$A self-energy in Fig.~\ref{fig2}\,(a) where the decrease of the weight of the pole $A$ corresponds to the increase of doping.
Fig.~\ref{fig2}\,(d) shows the optical conductivity $\sigma(\omega)$ and 
its difference to the Drude optical conductivity $\sigma(\omega) - \sigma_{Drude}(\omega)$ as the filled area with the same color. Hence, the latter represents the incoherent spectra originating from the pole structure. The MIR peak simultaneously shrinks and red-shifts as we decrease the weight $A$ of the PG pole. Eventually, as the pole structure vanishes, the MIR peak collapses into the Drude peak. Since the evolution of the MIR peak from this simplified model is consistent with Fig.~\ref{fig2}\,(b), we conclude that the MIR peak in the underdoped regime originates from the nodal gap structure generated by the same spin fluctuations which lead to the PG.

In comparison to the model $\sigma(\omega)$, the D$\Gamma$A result shows a more prominent MIR peak structure and a local maximum is observable until $\delta = 0.125$. This discrepancy originates from additional contribution from 
other momenta on the Fermi surface which have a PG but nearly zero or only a reduced quasiparticle weight.
This shifts the balance in favor of a more pronounced MIR peak, compared to the Drude peak. Supplementary section 3 elaborates for the details.

\textit{Mott-Hubbard correlations --} Aside from the (non-local) spin fluctuations, local Hubbard correlation lead to peaks in the optical spectrum. At half-filling, Hubbard bands develop around $\pm U/2$ in the one-particle spectrum, but  hole doping shifts the LHB to the Fermi energy where a QPB develops.
This leads to optical spectral weight in the MIR range and at slightly higher frequency around 1 eV ($2.5t$).

\begin{figure}
    \centering
    \includegraphics[width=0.96\linewidth]{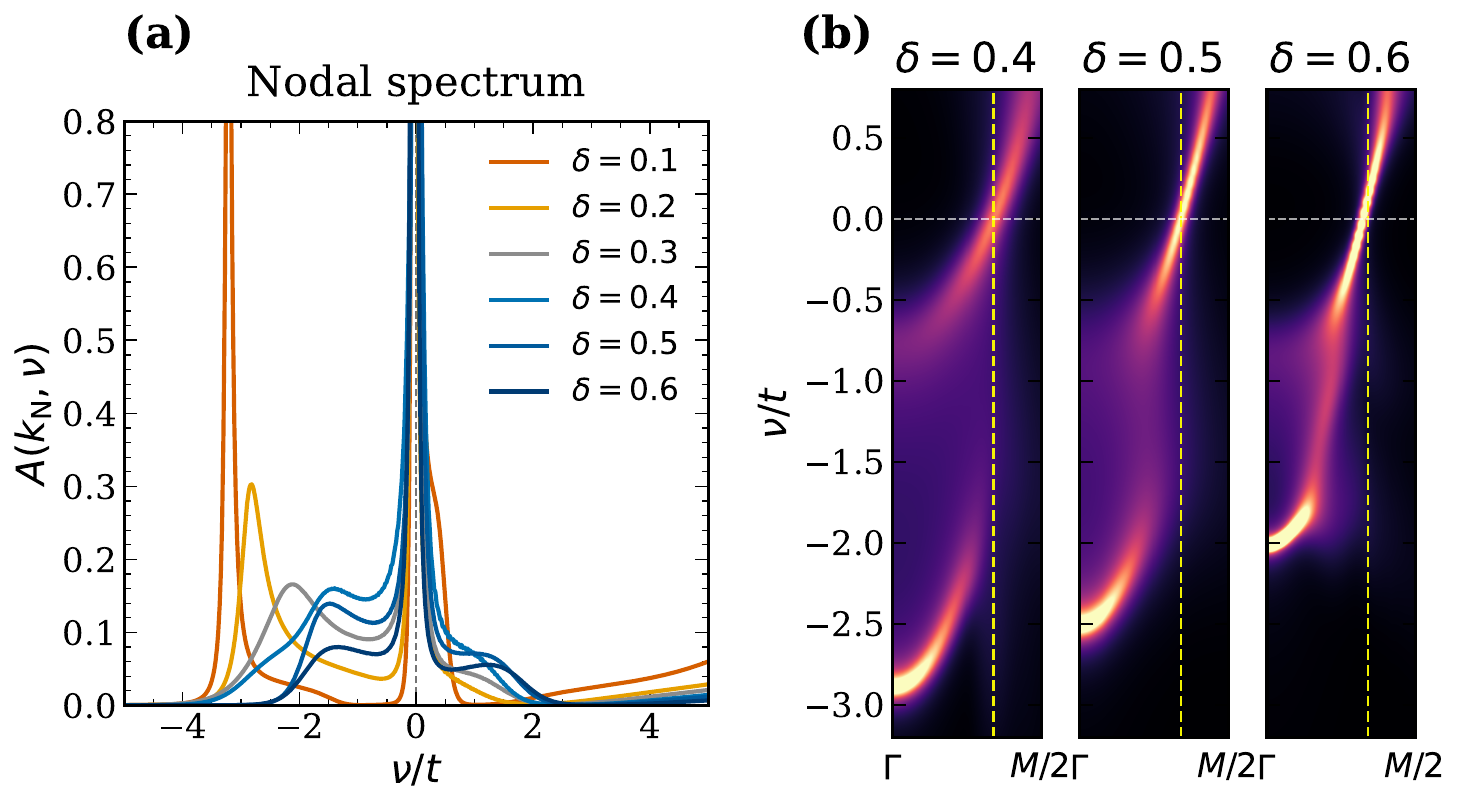}
    \caption{(a) Nodal spectrum $A(k_N, \nu)$ in a large frequency range for different dopings. (b) Doping dependence of  $A(k, \nu)$ along the $\Gamma-M/2$ line near the waterfall regime. The yellow vertical dashed lines denote the $k$-point corresponding to the nodal Fermi surface.}
    \label{fig3}
\end{figure}

\begin{figure}
    \centering
    \includegraphics[width=1.0\linewidth]{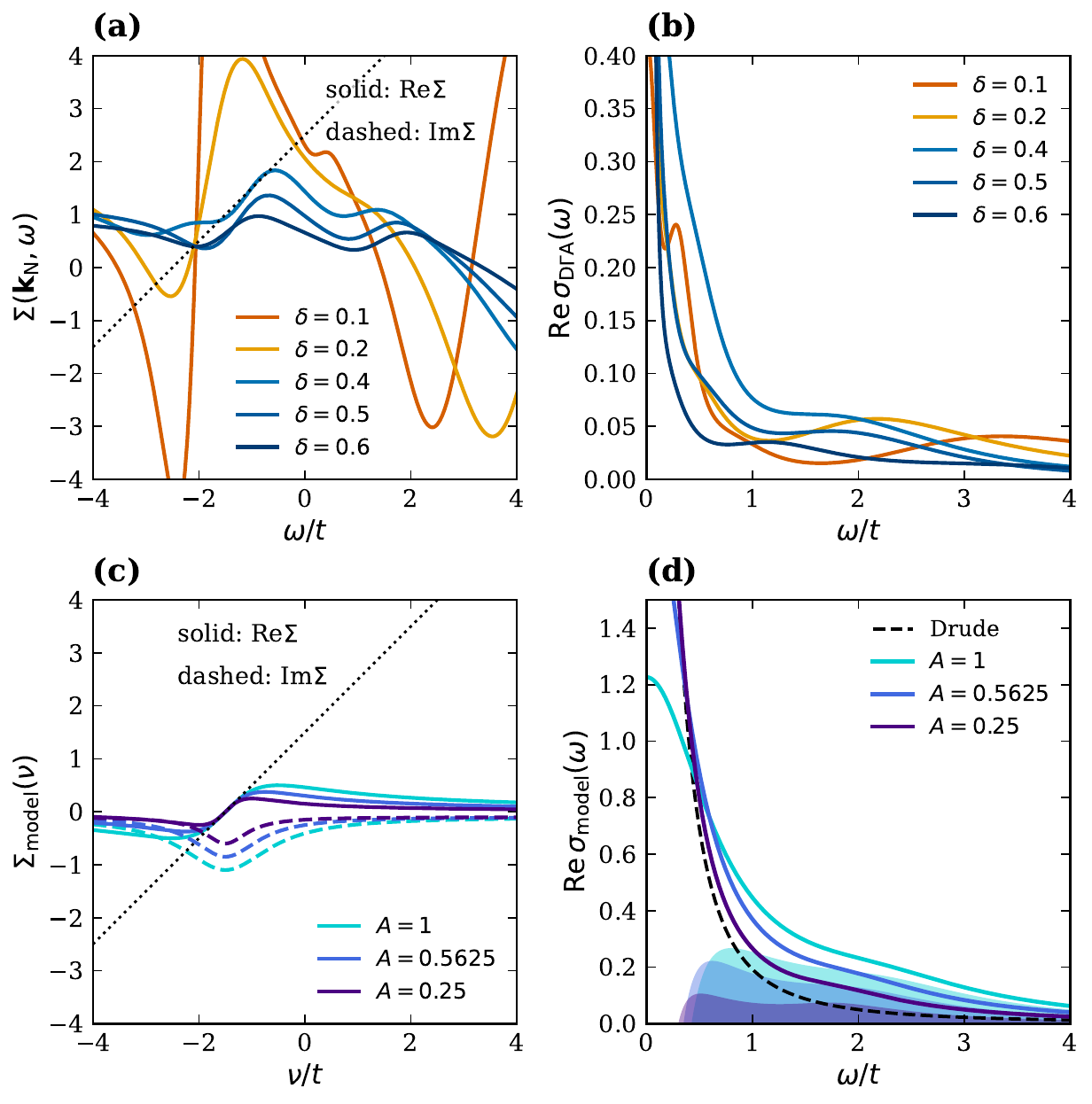}
    \caption{Evolution of (a) $\mathrm{Im}\Sigma(\omega)$ at the node and (b) the  optical conductivity with doping\khn{,} now on a large energy scale than in Fig.~\ref{fig3}. (c) Model self-energy with pole structure and $d \mathrm{Re}\Sigma(\nu)/d \nu\approx 1$ at the pole center to mimic the waterfall\khn. (d) Corresponding optical conductivity where the black dashed curve corresponds to the Drude optical conductivity $\sigma_{Drude}(\omega)$ and color fills represent the difference to the Drude optical conductivity, $\sigma(\omega) - \sigma_{Drude}(\omega)$.}
    \label{fig4}
\end{figure}

As shown in Fig.~\ref{fig3} (a), the LHB at the nodal FS shifts closer to the Fermi level as we increase the hole doping to $\delta = 0.4$. This generates a flat spectrum bridging %\khx{\textcolor{cyan}{between}} 
the QPB and LHB peaks,
%\khx{\textcolor{cyan}{illustrated by the} bridge-like nodal DOS}
and the origin of this flat spectrum (or filling of the gap between the peaks) is the waterfall discernible in Fig.~\ref{fig3}\,(b). The waterfall starts to develop at $\delta = 0.4$ and fully connects LHB and QPB at $\delta=0.6$. 
This waterfall originates from the 
the slope of the self-energy that separates QPB and  LHB. If the slope becomes one, i.e., $\mathrm d \mathrm{Re}\Sigma(\nu)/\mathrm d \nu = 1$ [black dashed line in Fig.~\ref{fig4}\,(a)], there are excitations for the same momentum but 
a range of frequencies, see
~\cite{krsnik2025local}. Thus, the slope of one is the origin of the flat spectrum below the Fermi level in Fig.~\ref{fig3} (a) and the waterfall in Fig.~\ref{fig3}\,(b).

Until a doping of $\delta = 0.4$, the LHB remains separated from the QPB but shifts closer to the Fermi energy with increasing doping in 
Fig.~\ref{fig3}\,(a). As a consequence, the optical spectra in Fig.~\ref{fig4}\,(b)  above $2t$ red shifts and gets closer to the MIR regime from $\delta=0.1$ to $\delta=0.4$. In contrast, from $\delta = 0.4$ onward, the waterfall regime leads to a plateau-shaped optical spectrum instead of a maximum. The magnitude of this plateau is reduced  with further doping, consistent with the evolution of the nodal spectrum. Hence, the Mott-Hubbard-driven optical spectra in the MIR regime has a flat shape and is weakened almost uniformly over the MIR energy window.
 At smaller dopings, QPB and LHB are well separated, resulting in a peaked structure now, given their larger separation, outside the MIR regime.

To understand the evolution of the broad optical spectra from the evolution of the softened LHB splitting near the waterfall regime, we again do a (semi-) analytical analysis. 
We use again  Eq.~(\ref{selfenergy_model}), now with $Z = 1$ and $\nu_g=-1.5$. Since  the real-part of the pole structure near $\nu = \nu_{g}$ simplifies to
\begin{equation}
\label{selfenergy_waterfall}
\frac{A(\nu - \nu_g)}{(\nu-\nu_g)^2 + \Delta^2} \approx \frac{A(\nu - \nu_g)}{\Delta^2} \; ,
\end{equation}
we can model the the waterfall with a slope of one for the self-energy by the constraint $A=\Delta^2$. The resulting self-energy is shown in Fig.~\ref{fig4}\,(c), which mimics Fig.~\ref{fig4}\,(a)
below the Fermi energy for dopings $\delta\geq 0.4$. 
 Decreasing $A$, while keeping $A/\Delta^2 = 1$ fixed, roughly simulates the evolution in Fig.~\ref{fig4}\,(a) between $\delta=0.4$ and 0.6.

 Fig.~\ref{fig4}\,(d) shows the optical conductivity for this model self-energy which indeed resembles Fig.~\ref{fig4}\,(b) for $\delta \geq 0.4$, i.e., in the MIR range, regarding both its shape and its doping dependence. In the model, thanks to the separability of the Drude contribution, the MIR (and higher frequency) spectra can be clearly identified from $\sigma(\omega) - \sigma_{Drude}(\omega)$ which is shown as the color fill. The evolution of MIR spectra is nearly frequency-independent, consistent to the result in Fig.~\ref{fig4} (b).
 This shows that the characteristic flat, nearly frequency-independent optical spectra which does not shift with doping, is a direct manifestation of the waterfall structure.

\textit{Further Discussions --} Let us conclude with the comparison to experiment: in $\mathrm{La_{2-x}Sr_{x}CuO_4}$ (LSCO)~\cite{uchida1991optical, kumar2023characterization}, the sharp peak formed at 0.5 eV for doping x = 0.02 ($=\delta$) red-shifts with increasing doping and merges into the Drude peak at x = 0.2, consistent with Fig.~\ref{fig2}\,(b) and (d). Broad spectra below 1 eV grow up to a doping of x=0.15, show no change  until x=0.2, and then, upon further doping to x=0.34, are suppressed in a frequency-independent way. This is consistent with the behavior shown in Fig.~\ref{fig4}\,(b) and (d). Also the well studied $\mathrm{YBa_{2}Sr_{3}CuO_{6+x}}$ (YBCO) shows the same qualitative behavior~\cite{cooper1993optical}: the peak below 0.5 eV red-shifts and gets merged at x=0.9, along with an overall increase of spectral weight below 1 eV. Spectral weight above 1 eV shifts to lower frequencies with doping, forming a broad structure.

%\textcolor{blue}{
%Finally we conclude the discussion with the comparison to infinite-layered Nickelate $\mathrm{Nd_{1-x}Sr_{x}NiO_2}$ (NSNO) superconductor~\cite{kim2026optical}, which is a Mott-Hubbard type material due to the weak hybridization between Ni and O.
%}

Given the consistency of 
our theoretical analysis  with experiment, we
can conclude that the optical conductivity in the MIR frequencies
originates from two physically different gaps:
magnetic fluctuations open a PG above the Fermi energy for the nodal momentum which is dominating the optical conductivity. This yields 
a pronounced peak in the optical conductivity which red-shifts upon further doping. Additionally 
the gap between lower Hubbard band and quasiparticle band
contributes in the MIR frequencies. In the underdoped regime this contribution is in form of a peak and still
in the frequencies of visible light. In the overdoped regime
these two bands connect through a waterfall, which leads to a flat, frequency-independent spectral contribution from the MIR to the NIR. Further doping uniformely supresses this frequency-independent spectrum. As the MIR optical conductivity is originated from two gaps, they can be complicatedly admixed in experiments. However, as we have shown, the contribution from both gaps can be distinguished by their characteristic shape and doping dependence.

\begin{acknowledgments}
{We acknowledge very helpful discussions with A.~Kauch, J.~Krsnik, J.~Peil }
This project has been supported by the Austrian Science Funds (FWF) through project DOI 10.55776/P36213.
\end{acknowledgments}

%\nocite{*}

\bibliographystyle{apsrev4-2}  % APS PRL citation style
\bibliography{references_opt}      % The name of your .bib file without the .bib extension

\clearpage                % 1. 본문의 잔여 텍스트와 그림을 완전히 밀어냄
\thispagestyle{empty}     % 2. 현재 페이지의 양식을 초기화
\cleardoublepage          % 3. 양면 인쇄 기준 다음 시작 페이지로 확실히 넘김

\renewcommand{\thefigure}{S\arabic{figure}}
\renewcommand{\theequation}{S\arabic{equation}}
\renewcommand{\thesection}{\arabic{section}}
\setcounter{figure}{0}
\setcounter{equation}{0}
\setcounter{section}{0}

\clearpage

\onecolumngrid

\begin{center}
{\Large\bfseries Supplemental Material for\\[0.3em]
\textit{``Two distinct gap structures in the mid-infrared
optical conductivity of the Hubbard model''}}\\[0.5em]

\vspace{1em}

Dongwook Kim\textsuperscript{1,*} and Karsten Held\textsuperscript{1,*}\\
\textit{Institute of Solid State Physics, TU Wien, 1040 Vienna, Austria}
\end{center}

\vspace{1em} 

\section{Details on Model and Methods}

\subsection{Square Lattice Hubbard model}

We study the single-orbital square-lattice Hubbard model with Hamiltonian
\begin{equation}
\label{eq0}
\begin{aligned}
H = -t\sum_{<ij>_{nn},\sigma}c^{\dagger}_{i,\sigma}c_{j,\sigma} -t'\sum_{<ij>_{nnn},\sigma}c^{\dagger}_{i,\sigma}c_{j,\sigma} \\ -t''\sum_{<ij>_{nnnn},\sigma}c^{\dagger}_{i,\sigma}c_{j,\sigma} - \mu \sum_{i,\sigma}n_{i, \sigma} + U\sum_{i}n_{i, \uparrow} n_{i, \downarrow}
\end{aligned}
\end{equation}
where $c^{\dagger}_{i,\sigma}$ ($c_{i,\sigma}$) is the creation (anhilation) operator for an electron of spin $\sigma = \uparrow, \downarrow$ at position $i$. Here nn, nnn and nnnn refer
to nearest-neighbor, next-nearest-neighbor and next-next-nearest-neighbor pais; the  corresponding hopping parameters that we employ are $t\equiv 1$, $t'=-0.25t$ and $t''=0.12t$ respectively; $\mu$ is the chemical potential, and $U$ the local repulsion here fixed to $U = 8.0t$; and $n_{i,\sigma}\equiv c^{\dagger}_{i,\sigma}c_{i,\sigma}$. The inverse temperature is $\beta = 60/t$ throughout the paper. This Hubbard model represents the $d_{x^2-y^2}$ orbital in high-temperature cuprate superconductors, where the typical hopping amplitude is $t=0.4\,$eV.

\subsection{Ladder Dynamical Vertex Approximation}

Ladder dynamical vertex approximation ($\mathrm{D}\Gamma\mathrm{A}$)~\cite{Toschi2007, worm2023} calculations are done based on local one- and two-particle Green’s functions that we calculate with continuous time quantum monte carlo (CTQMC)~\cite{gull2011continuous} using the  w2dynamics code~\cite{wallerberger2019w2dynamics} at the converged 
dynamical mean field theory (DMFT)~\cite{georges1996dynamical} solution. For the analytic continuation from Matsubara frequencies, the maximum entropy method as implemented in the \texttt{ana\_cont} package~\cite{Kaufmann2021} is used.

\subsection{Optical Conductivity Calculation}

Optical conductivity was calculated by the Kubo formula from the current-current bubble correlation function $\Pi_{xx}(i\omega)$ without vertex corrections as
\begin{equation}
\Pi_{xx}(i\omega_m)
= - \frac{e^2}{N_k \beta}
\sum_{k,\,\nu_n} |v_x(k)|^2 \;
G(k, i\nu_n)\,
G\big(k, i\nu_n + i\omega_m\big)
\label{eq_currentbubble}
\end{equation}
where $e$ denotes the electronic charge which we consider as 1, $\beta$ denotes the inverse temperature, $N_k$ denotes number of k-points used, $v_x(k)$ denotes Fermi velocity in $x$ direction and $G(k, i\nu_n)$, denotes the one-particle Green's function for momentum $k$ and Matsubara frequency $\nu_n$. Since the node has the largest bare Fermi velocity along the Fermi surface, the result mostly reflects the 1-particle spectra near the node. Without vertex corrections, the one-particle gap structure generated by strong correlation within a single band is fully responsible for the incoherent spectra in the optical conductivity. 
%\khc{this does not change for  inter-orbital (band) transitions, then it is just the bands between different bands in k-space}. 

\section{Drude optical conductivity for 1-dimensional rectangular bare DOS system and self-energy with only background scattering and renormalization}

If the self-energy is
\begin{equation}\notag
\Sigma(\nu) = -i\Gamma_0 + \bigg(1-\frac{1}{Z}\bigg)\nu
\end{equation}
where $\nu$ represents a real frequency, $\Gamma_0$  the background scattering rate and $Z$ the quasiparticle renormalization factor, the Green's function reads
\begin{equation}\notag
G(k,\nu) = \frac{Z}{\nu - Z\epsilon_k + iZ\Gamma_0} \;.
\end{equation}
Here, $k$ represents the momentum and $\epsilon_k$ the bare energy at momentum $k$. Then the spectral function is
\begin{equation}\notag
A(k,\nu) = -\frac{1}{\pi}\mathrm{Im}G(k,\nu) = \frac{1}{\pi}\frac{Z^2\Gamma_0}{(\nu - Z\epsilon_k)^2 + (Z\Gamma_0)^2}.
\end{equation}

The bubble part of optical conductivity for real frequencies reads
\begin{equation}\notag
\sigma(\omega) \propto \frac{1}{\omega} \int d\nu [f(\nu) - f(\nu + \omega)] \sum_k |v_k|^2 A(k, \nu)A(k, \nu+\omega)
\end{equation}
where $f(\nu)$ represents the Fermi-Dirac distribution function for frequency $\nu$, and $v_k$ is the Fermi velocity. For a constnat DOS adn $v_k$, it can be simplified to
\begin{equation}\notag
\sigma(\omega) \propto \frac{1}{\omega} \int d\nu [f(\nu) - f(\nu + \omega)] \int d\epsilon A(\epsilon, \nu)A(\epsilon, \nu+\omega)
\end{equation}
where the prefactors from $v_k$ and DOS are omitted. The $\epsilon$ integral is a  Lorentzian-Lorentzian convolution which yields
\begin{eqnarray}\notag
\int d\epsilon A(\epsilon, \nu)A(\epsilon, \nu+\omega)
&=& \frac{1}{\pi^2} \int d\epsilon \frac{\Gamma_0}{(\nu/Z - \epsilon)^2 + (\Gamma_0)^2} \frac{\Gamma_0}{((\nu + \omega)/Z - \epsilon)^2 + (\Gamma_0)^2} \\
&=& \frac{1}{\pi}\frac{2\Gamma_0}{(\omega/Z)^2 + (2\Gamma_0)^2}
\end{eqnarray}
which is independent of $\nu$. Thus, the remaining $\nu$ integral yields
\begin{equation}\notag
\sigma(\omega) \propto \frac{1}{\omega} \int d\nu [f(\nu) - f(\nu + \omega)] \approx 1
\end{equation}
for $\omega \gg T$. Thus, the resulting optical conductivity becomes
\begin{equation}\notag
\sigma(\omega) \propto \frac{Z}{\pi}\frac{2Z\Gamma_0}{\omega^2 + (2Z\Gamma_0)^2}
\end{equation}
which in a normalized version
\begin{equation}\notag
\sigma_{normalized}(\omega) = \frac{1}{\pi}\frac{2Z\Gamma_0}{\omega^2 + (2Z\Gamma_0)^2}.
\end{equation}
%\khc{Why use normalized version? Without a extra $Z$.} \dk{Optical sum rule fixes the normalization for the systems with same carrier concentration. In a realistic doping dependence of course the normalization varies, but for controlled analysis only considering the gap structure evolution I fixed it to controlled normalization.}
%\khc{This is only true if a larger (infinite) number of bands is included. For the Hubbard model the sum rule fixes it to the kinetic energy (if $t'=t''=0$, otherwise a modification of it).}
For the black dashed Drude optical conductivity in Fig. 3 (d) and Fig. 5. (d), normalized Lorentzian was used, along with the same normalization for semianalytical results.

%$\frac{Z}{\pi}\frac{2Z\Gamma_0}{\omega^2 + (2Z\Gamma_0)^2}$ is used not $\sigma_{normalized}(\omega)$ although in the Eq. 4 we showed the normalized version for the clear illustration of Lorentzian.

\section{Model analysis on the contribution from intermediate region between node and antinode}

\begin{figure}
    \centering
    \includegraphics[width=0.96\linewidth]{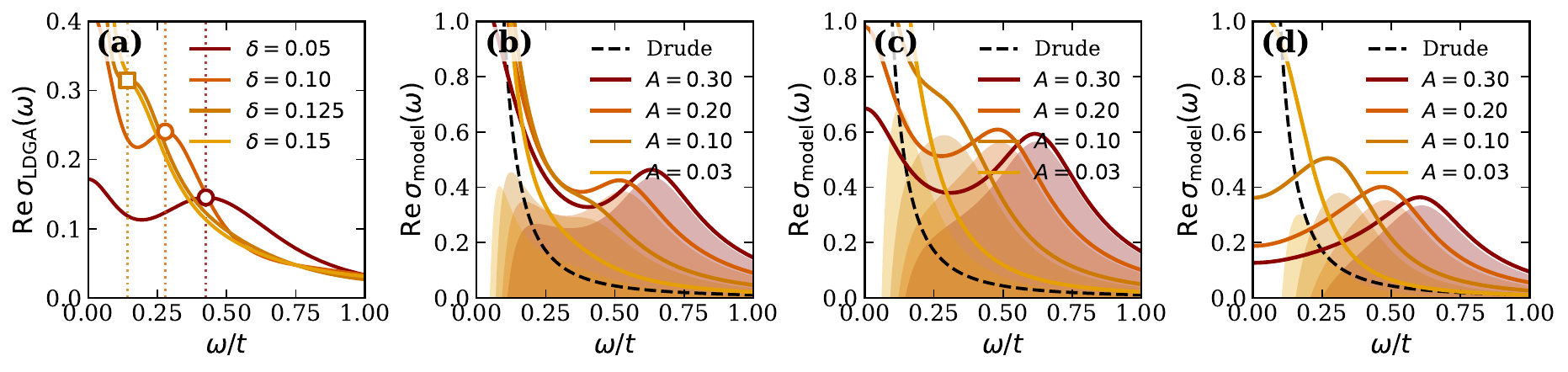}
    \caption{(a) Evolution of optical conductivity as dopings from D$\Gamma$A. Dotted vertical lines denote the peak frequency, which correspond to 0.42, 0.28 and 0.14 for $\delta$ of 0.05, 0.10 and 0.125 respectively. Evolution of optical conductivity for different weights $A$ of the pole in the model self-energy, with (b) a pole centered at $\omega=0.2t$ %Fig. 3 (c)) 
    and (d) a pole centered at $\omega = 0$.
    (c) Average of (b) and (d), which mimics the mixed contribution between node and intermediate region of the Fermi surface.
    Color fills represent the respective difference to the Drude optical conductivity, $\sigma(\omega) - \sigma_{Drude}(\omega)$.}
    \label{figs1}
\end{figure}

As shown in Fig. 2 (e) and (f) of the main text, the intermediate region of the Fermi surface between node and antinode, which has a smaller Fermi velocity than the node but still a finite one in contrast to the antinode, has a gap structure near the Fermi level, which leads to the MIR peak without Drude peak. Thus these points can lead to more prominent MIR peak structure in the D$\Gamma$A result than the analytical model result which describes the nodal contribution only. Fig.~\ref{figs1} compares D$\Gamma$A result in Fig.~\ref{figs1} (a) with analytical model self energy results. Here, Fig.~\ref{figs1} (b) shows the optical conductivity from the self-energy with pole structure located at $\omega = 0.2t$ (i.e., the model for the nodal contribution), which is also shown in Fig. 3 (b) of the main text. 
Fig.~3 (d) instead presents the optical contribution for intermediate momenta with a model self-energy that has a pole and thus a gap at the Fermi energy; Fig.~3 (c) is the averaged contribution of (b) and (d) which now has a discernible extra peak next to the Drude peak as in the D$\Gamma$A result of panel (a).

\end{document}